\documentclass{article}
\usepackage{ismir,amsmath,cite}
\usepackage{graphicx}
\usepackage{color}
\usepackage[utf8]{inputenc}

\title{Live Score Following on Sheet Music Images}

\multauthor
{Matthias Dorfer$^1$ \hspace{1cm} Andreas Arzt$^2$ \hspace{1cm} Sebastian Böck$^1$} { \bfseries{Amaury Durand$^3$ \hspace{1cm} Gerhard Widmer$^{1,2}$}\\
$^1$ Department of Computational Perception, Johannes Kepler University, Austria\\
$^2$ Austrian Research Institute for Artificial Intelligence, Linz, Austria\\
$^3$ Télécom ParisTech\\
{\tt\small matthias.dorfer@jku.at}
}
\def\authorname{Matthias Dorfer, Andreas Arzt, Sebastian Böck, Amaury Durand, Gerhard Widmer}

\begin{document}
\maketitle
\begin{abstract}

In this demo we show a novel approach to score following. Instead of
relying on some symbolic representation, we are using a multi-modal
convolutional neural network to match the incoming audio stream directly to
sheet music images. This approach is in an early stage and should be seen as
proof of concept. Nonetheless, the audience will have the opportunity to test
our implementation themselves via 3 simple piano pieces.

\end{abstract}

\section{Introduction}\label{sec:introduction}

Commonly, score following is defined as the process of following a musical
performance (audio) with respect to some representation of the sheet music.
State-of-the-art algorithms can e.g. be found in
\cite{Cont_2009_CDFA,Raphael_2010_MusicPlusOne,Prockup_2013_Orchestra,Arzt_2015_AIConcertgebouw}.
All of these approaches depend on some symbolic representation of the sheet
music, to which the incoming audio stream is matched. In this demo we present
the first approach that is able to operate directly on sheet music images. The
presented system still is at a very early stage and is only capable of following
very simple music (monophonic, notated on a single staff). Nonetheless, we
believe that this is a very promising approach and hope that it will spark
further research.

\section{Score Following on Sheet Music Images}\label{sec:score_following}

\begin{figure*}
 \centerline{
 \includegraphics[width=15cm]{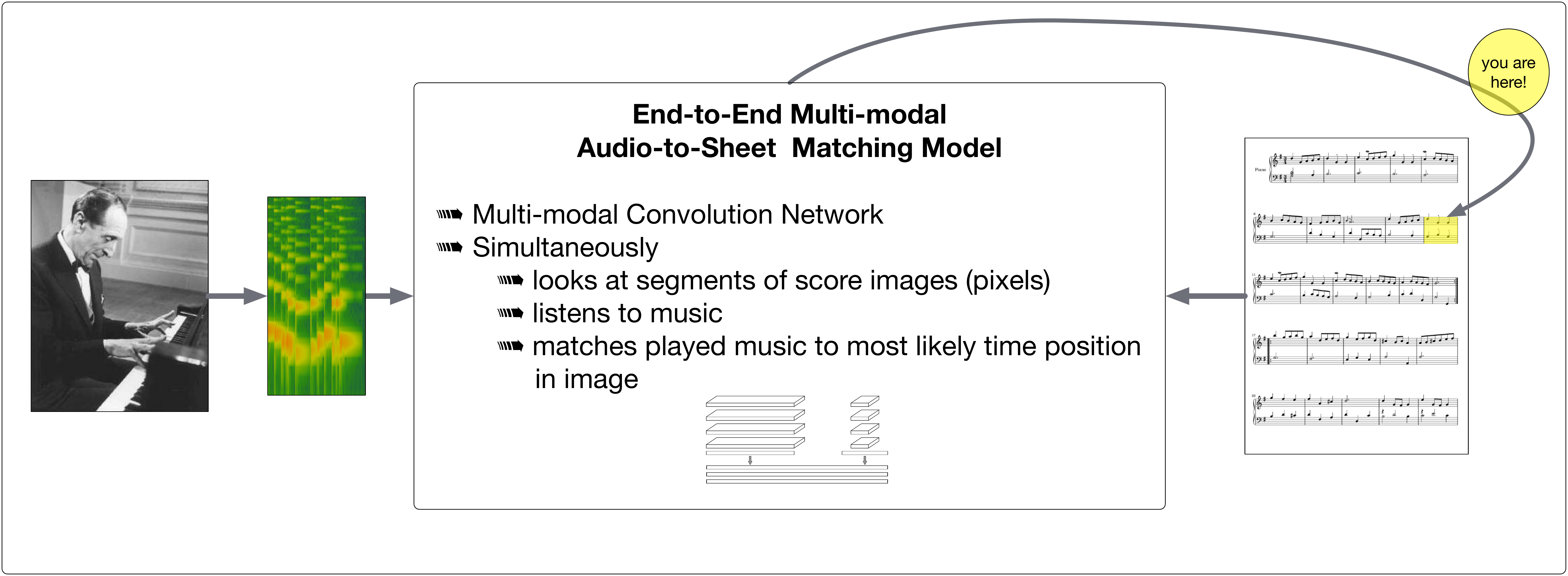}}
 \caption{Task overview. A multi-modal convolution network is taking a live audio stream as input, and computing the most probable position in the sheet music.}
 \label{fig:task}
\end{figure*}

Figure \ref{fig:task} shows an overview of the setup. The audio signal of the live
performance is captured via a microphone, after which we apply some light
preprocessing. We compute log-spectrograms with an audio sample rate of
22.05kHz, FFT window size of 2048 samples, and computation rate of 15 frames
per second. To reduce the dimensionality, we apply a normalised 24-band
logarithmic filterbank, allowing only frequencies from 80Hz to 8kHz. This
results in 136 frequency bins. The audio processing part is done with an on-line
capable version of the \emph{madmom} library\cite{Boeck_2016_Madmom}.

A context of 40 frames -- roughly 2.7 seconds -- is provided to a multi-modal
convolutional neural network (see \cite{Dorfer_2016_Towards} for details on the
architecture and the training process, as well as off-line recognition results).
This network has been trained to match audio in various different tempi to
(parts of) a sheet music image. We are using a context of exactly one staff,
centred at the previously detected position, and linearly quantised into 40
bins. The network computes the probability of a match between the audio context
and each of the 40 bins, and returns the most probable one as the current
position in the score.

So far our approach does not keep any history of the tracking process (except
for the previous position). This means that a few misclassifications of the
network might lead to big jumps in the score, with the effect that the tracker
is getting lost. While a natural approach would be to incorporate the tracking
history into the model itself, we for now opted for a simpler approach and added
an optional post-processing step via an on-line Dynamic Time Warping algorithm
to smooth the output of the network and thus stabilise the tracking process.

\section{The Live Demo}\label{sec:demo}

Due to hardware limitations -- the demonstration will be shown on a common
laptop with no graphics card present that would be able to provide the
computational power typically needed for deep learning algorithms -- we will not
be able to show the full capabilities of our approach.

Most importantly, we had to train our model directly on the pieces in question,
as more general models tended to get too large to be decoded in real-time on the
available hardware. However, note that for the non-real-time (but strictly
left-to-right) scenario described in \cite{Dorfer_2016_Towards} we have already
successfully trained models that can cope with new (unseen) pieces and would
lead to comparably good tracking results. We hope to be able to demonstrate this
live in the very near future.


For the very same reason, we will only demonstrate score following on monophonic
music, notated on a single staff. We already did preliminary experiments which
suggest that our approach is easily generalisable to more complex, polyphonic
music, notated on multiple staves. Again, we hope to be able to demonstrate this
to the public at a later stage.

\begin{figure}
 \centerline{
 \includegraphics[width=\columnwidth]{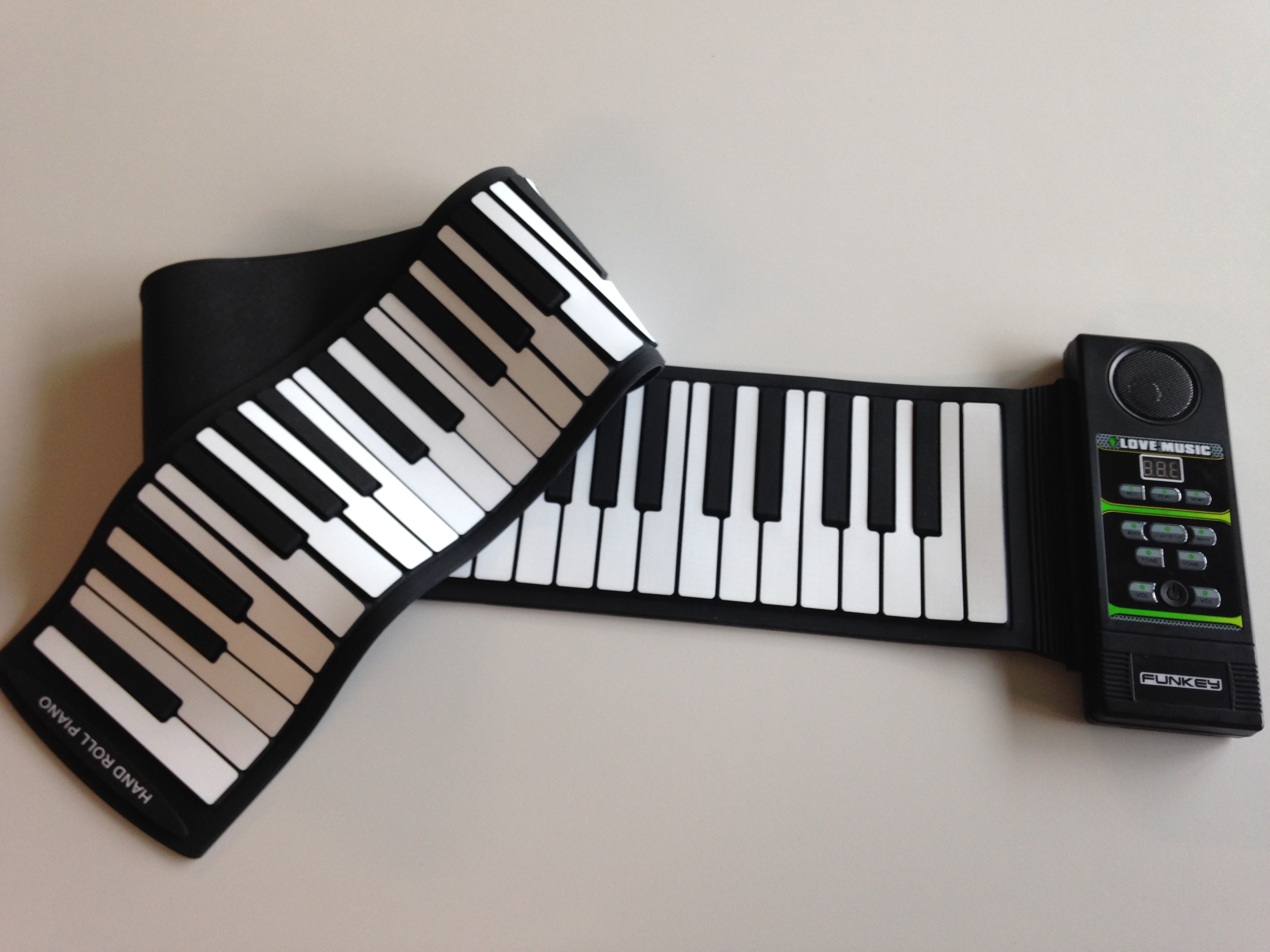}}
 \caption{The portable piano that we will use for the demo.}
 \label{fig:piano}
\end{figure}

For the demo, we prepared excerpts of 3 pieces: the lullaby \emph{Twinkle
Twinkle Little Star}, Bach’s \emph{Minuet in G-major} and Gigi d’Agostino’s
\emph{The Riddle}, due to certain similarities between the visualisation of the
output of the neural network and the cartoon character La Linea, which was used
in his music video. For all 3 pieces we only prepared the monophonic melody,
with no accompaniment. The audience is very welcome to try to play these pieces
on our portable piano keyboard (see Figure \ref{fig:piano}) and test our algorithm.

\section{Conclusions}\label{sec:conclusion}

This demonstration shows very early work on score following directly on sheet
music and should be seen as a proof of concept. Future work includes lifting the
limitations of our approach, with the first steps being to train models that can
read more complex, polyphonic music, notated on multiple staffs.

\section{Acknowledgements}

This work is supported by the Austrian Ministries BMVIT and BMWFW,
and the Province of Upper Austria via the COMET Center SCCH,
and by the European Re\-search Coun\-cil (ERC Grant Agreement
670035, project CON ESPRESSIONE; FP7 grant agreement no. 610591, project GIANT\-STEPS).
The Tesla K40 used for this research was donated by the NVIDIA
corporation.

\bibliography{ISMIR2016-LBD}

\end{document}